\documentclass[12pt]{article}

\usepackage{graphics}
\usepackage{epsfig}
\usepackage{mathrsfs}

\newcommand{\nn}{\nonumber}

\newcommand{\bq}{\begin{eqnarray} }
\newcommand{\eq}{\end{eqnarray} }
\newcommand{\e}{\epsilon}

\begin{document}
\begin{titlepage}
\begin{flushright}
OSU-HEP-05-01\\
January 2005\\
\end{flushright}

\vspace*{1.5cm}
\begin{center}
{\Large {\bf Split Supersymmetry from Anomalous $U(1)$\\[-0.05in]
} }

\vspace*{1.5cm}
 {\large {\bf K.S. Babu\footnote{E-mail address:
 babu@okstate.edu}, Ts. Enkhbat\footnote{E-mail address:
 enkhbat@okstate.edu}}}\\
{\it Department of Physics, Oklahoma State University\\
Stillwater, OK~74078, USA }\\and\\ {\large{\bf Biswarup
Mukhopadhyaya\footnote{E-mail address: biswarup@mri.ernet.in}
 }}\\

{\it Harish-Chandra Research Institute,\\ Chhatnag Road, Jhusi,
Allahabad - 211 019, India }
\end{center}

 \vspace*{1.5cm}

\begin{abstract}

We present a scenario wherein the anomalous $U(1)$ ~$D$--term of
string origin triggers supersymmetry breaking and  generates
naturally a Split Supersymmetry spectrum. When the gaugino and the
Higgsino masses (which are of the same order of magnitude) are set
at the TeV scale, we find the scalar masses to be in the range
$(10^6 - 10^8)$ GeV. The $U(1)$ ~$D$--term provides a small
expansion parameter which we use to explain the mass and mixing
hierarchies of quarks and leptons. Explicit models utilizing exact
results of $N=1$ suersymmetric gauge theories consistent with
anomaly constraints, fermion mass hierarchy, and supersymmetry
breaking are presented.

\end{abstract}

\end{titlepage}
\newpage

\section{Introduction}

~~~~~It is widely believed that supersymmetry may be relevant to
Nature.  There are four major observations which may justify this
belief: (i) Supersymmetry (SUSY) can stabilize scales associated
with spontaneous symmetry breaking. (ii) Unification of gauge
couplings works well in the minimal SUSY extension of the Standard
Model (SM). (iii) SUSY provides a natural candidate for cold dark
matter. (iv) Supersymmetry is a necessary ingredient of
superstring theory, which may eventually lead to a consistent
quantum theory of gravity.  Among these, reasoning (i), when
applied to stabilize the electroweak scale, would suggest that all
superpartners  of the SM particles must have masses below or
around a TeV. This is indeed what was assumed in almost all
applications of supersymmetry to particle physics in the past
twenty five years. The second and third observations above would
only require that a subset of  superpartners be lighter than a
TeV, while the last one allows SUSY to be broken anywhere below
the Planck scale, $M_{Pl} = 2.4 \times 10^{18}$ GeV. This is
because, among the superpartners, if the split members of a
unifying group ($SU(5)$, $SO(10)$, etc), namely the gauginos and
the Higgsinos, are lighter than a TeV, while the complete
multiplets (the scalar partners of SM fermions) are much heavier,
unification of gauge couplings would work just as well. The
lightest of these SUSY particles would still be a natural
candidate for cold dark matter.

A scenario  dubbed as ``Split Supersymmetry'',  in which the spin
1/2 superparticles, namely, the gauginos and the Higgsinos, have
masses of order TeV while the spin zero superparticles (squarks
and sleptons) are much heavier, has recently been advocated
\cite{Arkani-Hamed:2004fb}. This scenario gives up the
conventionally employed naturalness criterion, since the light SM
Higgs boson is realized only by fine--tuning.  Such a finely tuned
scenario, it is argued, may not be as improbable as originally
thought \cite{Arkani-Hamed:2004fb}. This is because in any theory
with broken SUSY one has to cope with another, even more severe,
fine-tuning, in the value of the cosmological constant.  A cosmic
selection rule, an anthropic principle \cite{Weinberg:1987dv}, may
be active in this case. If so, a similar argument may also explain
why the SM Higgs boson is light \cite{Agrawal:1998xa}.
Supersymmetry plays no role in solving the hierarchy problem here.
Recent realization of a string landscape \cite{Bousso:2000xa},
which suggests the existence of a multitude of string vacua, may
justify this approach. Probabilistically, the chances of finding a
vacuum with a light SM Higgs (along with a small cosmological
constant) may not be infinitesimal, given the existence of a large
number of string vacua \cite{Banks:2003es}.

Split Supersymmetry has a manifest advantage over TeV scale
supersymmetry: Unacceptably large flavor changing neutral current
(FCNC) processes \cite{Wells:2004di}, fermion electric dipole
moments, and $d=5$ proton decay rate, which generically plague TeV
scale SUSY are automatically absent in Split Supersymmetry.
Various aspects of this scenario have been analyzed by a number of
authors \cite{gr,other}.

In this paper we take the Split Supersymmetry scenario from a
theoretical point of view. Perhaps the most important question in
this context is a natural realization of the split spectrum.
Although it may be argued that $R$--symmetries would protect
masses of the spin 1/2 SUSY fermions and not of the squarks and
sleptons, in any specific scenario for SUSY breaking there is very
little freedom in choosing the relative magnitudes of the two
masses.  We will focus on SUSY breaking triggered by the anomalous
$U(1)$ ~$D$--term of string origin  coupled to a SUSY QCD sector
\cite{Binetruy:1996uv}. Each sector treated separately would
preserve supersymmetry, but their cross coupling breaks it. We
make extensive use of exact results known for $N=1$ SUSY QCD
\cite{Seiberg:1994bz}. In this scenario, the squarks and sleptons
receive SUSY breaking masses at the leading order from the
anomalous $U(1)$ ~$D$--term, while the gauginos acquire masses
only at higher order.  The Higgsino mass also arises at higher
order and is similar in magnitude to the gaugino mass. Thus, a
naturally split spectrum is realized. The anomalous
$U(1)$~$D$--term also provides a small expansion parameter which
we use to explain the mass and mixing hierarchies of quarks and
leptons.  We present complete models which are consistent with
anomaly cancelation, and which lead to naturally split SUSY
spectrum.\footnote{A somewhat similar analysis has recently been
carried out in Ref. \cite{nath}, our approach is different in that
we present complete models without assuming a hidden sector and
address the fermion masses and mixing hierarchy problems. Our
spectrum is also quite different, especially as regards the
gravitino mass.} We note that with flavor--dependent charges, the
anomalous $U(1)$ $D$--term contributions to the squark and slepton
masses generically lead to large FCNC processes with sub--TeV
scalars \cite{Kawamura:1994ys}, this problem is absent in the
Split Supersymmetry scenario.

\section{Supersymmetry breaking by anomalous $U(1)$ and gaugino condensation}

~~~~~In this section we review supersymmetry breaking induced by
the $D$--term of anomalous $U(1)$ symmetry
\cite{Binetruy:1996uv,pomarol} coupled to the strong dynamics of
$N=1$ SUSY gauge theory \cite{Seiberg:1994bz}.   Each sector
separately preserves supersymmetry, so an expansion parameter (the
cross coupling) is available. Exact results of supersymmetric
gauge theories can then be applied. Here we focus on the global
supersymmetric limit, in Sec. 2.1 we extend the analysis to
supergravity.  In addition to the SM fields, these models contain
an $SU(N_c)$ gauge sector with $N_f$ flavors. The ``quark"  ($Q$)
and ``antiquark" ($\tilde{Q}$) fields of the $SU(N_c)$ sector are
also charged under the $U(1)_A$. $U(1)_A$ is broken by a SM
singlet field $S$ carrying $U(1)_A$ charge of $-1$. The Standard
Model fields carry flavor--dependent  $U(1)_A$ charges so that the
hierarchy in fermion masses and mixings is naturally explained.  A
small expansion parameter $\epsilon \sim 0.2$ is provided by the
ratio $\epsilon = \left\langle S \right \rangle/M_{Pl}$ by the
induced Fayet--Iliopoulos $D$--term for the $U(1)$.  To see this,
we recall that the apparent anomalies in $U(1)_A$ are canceled by
the Green--Schwarz (GS) mechanism \cite{gs}.  Heterotic
superstring theory when compactified to four dimensions contains
the Lagrangian terms $L \supset \varphi(x) \sum_ik_i F_i^2 +
i\eta(x) \sum_ik_i F_i \tilde{F}_i$, where $k_i$ are the
Kac--Moody levels, $\varphi(x)$ is the dilaton field and $\eta(x)$
is its axionic partner.  The GS mechanism makes use of the
transformation $\eta(x) \rightarrow \eta(x) - \theta(x)
\delta_{GS}$, and the gauge variation for the $U(1)_A$ gauge
field, $V_\mu \rightarrow V_\mu + \partial_\mu \theta(x)$.  The
anomalies are canceled if the following conditions are satisfied:
\bq\label{anomaly1}\frac{A_{i}}{k_i}=\frac{A_{N}}{k_N}
=\frac{A_{A}}{3k_A}=\frac{A_{{\rm gravity}}}{24} = \delta_{GS},\eq
where $A_i\,(i=1,2,3)$, $A_N$, $A_A$ and $A_{\rm gravity}$ are the
anomaly coefficients for $SM^2\times U(1)_A$, $SU(N_c)^2\times
U(1)_A$, $U(1)_A^3$ and gravity$^2\times U(1)_A$.  Here $A_{\rm
gravity}$ is the gravitational anomaly, given by the sum of the
anomalous charges of all fields in the theory. All other anomalies
must vanish. These conditions put severe restrictions on the
choice of $U(1)_A$ charges.

String loop effects induce a nonzero  Fayet--Iliopoulos $D$--term
for the $U(1)_A$ given by \cite{Dine:1987xk, Atick:1987gy}
\begin{eqnarray}\label{FIterm}
\xi=\frac{g_{st}^2M_{Pl}^2}{192\pi^2}A_{\rm gravity}\,,
\end{eqnarray}
where $g_{st}$ is the string coupling at the unification scale
$M_{Pl}$, related to the SM gauge couplings at that scale as
\begin{equation} k_i g_i^2 = 2 g_{st}^2~.
\end{equation}
The scalar potential receives a contribution from the $D$-term
given by
\begin{eqnarray}\label{VD0}
V_D={D_A^2 \over 2}=\frac{g_A^2}{2}\left(\xi-|S|^2+q_{Q}|Q_i|^2+
q_{\tilde{Q}}|\tilde{Q}^{\tilde{i}}|^2+\sum_iq_i|\phi_i|^2\right)^2.
\end{eqnarray}
Here $S$ is the flavon field with charge $-1$, $Q_i$ and
$\tilde{Q}^{\tilde{i}}$ are the ``quark" and ``antiquark" fields
belonging to the fundamental and antifundamental representaions of
an $SU(N_c)$ gauge group with $U(1)$ charges $q_Q$ and
$q_{\tilde{Q}}$.  $\phi_i$ in Eq. (4) stand for all the other
fields, and includes the SM sector.

In our models, all fields except $S$, will have positive $U(1)_A$
charges, so $\xi$ will turn out to be positive.  The potential of
Eq. (4) will minimize to  preserve supersymmetry by giving the
negatively charged $S$ field a vacuum expectation value (VEV),
which would break the $U(1)_A$ symmetry. To zeroth order in SUSY
breaking parameters, $\left\langle S \right\rangle = S_0$, where
\begin{eqnarray}\label{vevS0}
S_0\equiv\sqrt{\xi}=\sqrt{\frac{g_{st}^2A_{\rm
gravity}}{192\pi^2}}M_{Pl}\equiv \e M_{Pl}.
\end{eqnarray}
Here $\e \sim 0.2$ will provides a small expansion parameter to
explain the hierarchy of quark and lepton masses and mixings.

As for the $N=1$ SUSY QCD sector, we consider the gauge group
$SU(N_c)$ with $N_f$ flavors of quarks and antiquarks, and  apply
the well--known exact results of Ref. \cite{Seiberg:1994bz}. For
concreteness we choose $N_f < N_c$.  These results have been
applied to TeV scale SUSY breaking by Binetruy and Dudas in Ref.
\cite{Binetruy:1996uv} in the presence of anomalous $U(1)$
symmetry.  These models actually lead to a Split Supersymmetry
spectrum, as we will show.  We also generalize the results of Ref.
\cite{Binetruy:1996uv} to include supergravity corrections (in
Sec. 2.2).  In Sec. 3, we apply these results to explicit and
complete models.

The effective superpotential we consider has two pieces:
\begin{equation}
W_{\rm eff} = W_{\rm tree} + W_{\rm dynamical},
\end{equation}
where $W_{\rm tree}$ is the tree--level superpotential, while
$W_{\rm dynamical}$ is induced dynamically by nonperturbative
effects.  Since the $Q$ and the $\tilde{Q}$ fields are charged
under $U(1)_A$, a bare mass term connecting them is not allowed. A
mass term will arise through the coupling
\begin{equation}
W_{\rm tree} =\frac{\mbox{Tr}\left(\lambda
Q\tilde{Q}\right)S^n}{M_{*}^{n-1}}
\end{equation}
when $\left \langle S \right \rangle = S_0$ is inserted. Here the
trace is taken over the $N_f$ flavor indices of the $Q_i$ and
$\tilde{Q}^{\tilde{i}}$ fields. $M_*$ is a mass scale at which
this term is induced.  The most natural value of $M_*$ is
$M_{Pl}$, which is what we will use for our numerical analysis,
but we allow $M_*$ to be different from $M_{Pl}$ for generality.
We have used the definition
\bq\label{defn}n=q_{Q}+q_{\tilde{Q}}\eq for the sum of the $U(1)$
charges of $Q$ and $\tilde{Q}$. As we will see later the choice
$n=1$, which would correspond to a  renormalizable superpotential
will be  phenomenologically unacceptable.  From the results of
Ref. \cite{Seiberg:1994bz}, the dynamically generated
superpotential is known to be (for $N_f < N_c)$
\begin{equation}\label{superP}
W_{\rm dynamical} = (N_c-N_f)\left(\frac{\Lambda^{3N_c-N_f}}
{\mbox{det}\left(Q\tilde{Q}\right)}\right)^{1/(N_c-N_f)}.
\end{equation}
Here  $\Lambda$ is the dynamically induced scale below which the
$SU(N_c)$ sector becomes strongly interacting:
\bq\label{defLambda}\Lambda\sim
M_{Pl}e^{-\frac{2\pi}{\alpha_{N_c}(3N_c-N_f)}},\eq where
$\alpha_{N_c}$ is the $SU(N_c)$ gauge coupling constant at
$M_{Pl}$. For $N_f = N_c-1$, the gauge symmetry is completely
broken, and Eq. (\ref{superP}) is induced by instantons.  For $N_f
< N_c-1$, the gauge symmetry is reduced to $SU(N_c-N_f)$ and the
gaugino condensate of this symmetry induces Eq. (\ref{superP}).

Below the scale $\Lambda$ the effective theory can be described in
terms of $N_f \times N_f$ mesons $Z^{\tilde{i}}_{j}$:
\bq\label{defZ}Z^{\tilde{i}}_{j}=
Q_j\tilde{Q}^{\tilde{i}}\,\,\,\,\mbox{with}\,\,\,\,(\tilde{i},j=1,..,N_f).\eq
Neglecting small supersymmetry breaking effects, we can describe
the theory below $\Lambda$ along the $D$--flat directions $Q_i =
\tilde{Q}_i$ in terms of the $Z$ fields.  We can make the
following replacements in the $D$--term and the superpotential:
$q_{Q}|Q_i|^2+q_{\tilde{Q}}|\tilde{Q}^{\tilde{i}}|^2\rightarrow
n\mbox{Tr}\left(Z^\dagger Z\right)^{1/2}$ and $
Q_j\tilde{Q}^{\tilde{i}}\rightarrow Z^{\tilde{i}}_{j}$.  We use
the notation \bq\label{defm}m=\lambda\frac{S_0^n}{M_{*}^{n-1}},\eq
with $m$ identified as the mass matrix of the $Z$ field (upto
small supersymmetry breaking effects). Then the $F$--term for the
$Z$ fields, defined as $\left(
F_Z\right)^{\tilde{i}}_j=2\left[\left(Z^\dagger
Z\right)^{1/2}\right]^{\tilde{i}\tilde{k}}\partial W/\partial
Z_{j}^{\tilde{k}}$, is found to be
\bq\label{Fz0}\left(F_Z\right)^{\tilde{i}}_j=2\left[\left(Z^\dagger
Z\right)^{1/2} \left(m
-\left(\frac{\Lambda^{3N_c-N_f}}{\mbox{det}\left(Z\right)}\right)^{1/(N_c-N_f)}
\left(\frac{1}{Z}\right)\right)^T\right]^{\tilde{i}}_j.\eq This
theory preserves supersymmetry, as  $F_{Z}=0$ can be realized with
$\left \langle Z \right \rangle\neq 0$ and given by
\bq\label{vevZ0} \left(Z_0\right)^{\tilde{i}}_j\equiv
\left(\mbox{det}\left(m\right)
\Lambda^{3N_c-N_f}\right)^{1/N_c}\left(\frac{1}{m}\right)^{\tilde{i}}_j.\eq
Note that this result holds only in the presence of a nonvanishing
VEV $\left \langle S \right \rangle$, so that $m$ is nonzero.

So far we treated the $U(1)_A$~ $D$--term and the ensuing
superpotential for the $Z$ fields separately. The two sectors are
however coupled through $W_{\rm tree}$ of Eq. (7).  Owing to this
coupling, supersymmetry is actually broken. This is evident by
examining the $F$--term  of the $S$ field, \bq F_{S}=n \frac{Tr(m
Z_0) }{S_0} \neq 0.\eq Similarly $F_Z$ is also nonzero. The VEVs
of $S$ and $Z$ fields will shift from the supersymmetry preserving
values of Eqs. (5) and (14) when the full potential is minimized
jointly. To find the soft SUSY breaking parameters we need to
calculate these corrections.

The scalar potential of the model in the global limit is given by
\begin{eqnarray}\label{scalarPotential0}
V=|F_S|^2+\frac{1}{2}\mbox{Tr}(F_Z(Z^\dagger Z)^{-1/2}F^\dagger_Z)
+\frac{1}{2}D_A^2.
\end{eqnarray}
We expand the fields around the SUSY preserving minima:
\bq\label{delta0}S=S_0+\delta S\,\,\,\,\,\,\,\,Z^{\tilde{i}}_j
=\left(Z_0+\delta Z\right)^{\tilde{i}}_j\eq with $\delta S/S_0 \ll
1, ~\delta Z/Z_0 \ll 1$.  For simplicity we assume the coupling
matrix $\lambda$ to be an identity matrix, $\lambda^i_j = \lambda
\delta^i_j$, in which case $Z_i^j = Z \delta_i^j$ can be chosen.
The VEV $\left\langle Z \right \rangle = Z_0$ arising from Eq.
(14) in this case becomes
\begin{eqnarray}\label{Zvev}
Z_0=\frac{\Lambda^3}{m}\left(\frac{m}{\Lambda}\right)^{N_f/N_c}.
\end{eqnarray}
We make an expansion in the supersymmetry breaking parameter
$\Delta$ defined as
\begin{eqnarray}\label{Deltavev}
\Delta\equiv Z_0/ S_0^2=\frac{\Lambda^3}{m
S^2_0}\left(\frac{m}{\Lambda}\right)^{N_f/N_c}\ll 1.
\end{eqnarray}

From the minimization of the scalar potential with respect to
these shifted fields, we find \bq\label{deltas}\langle S^\dagger
S\rangle&=&S^2_0\left[1+\Delta
\left(nN_f\right)-\Delta^2\left(\frac{n^2N_f^2}{2N_c^2g_A^2}\right)\left\{\frac{}{}^{}
g_A^2n\left( N_c-N_f\right)\left(
2N_c-N_f\right)\right.\right.\nn\\&-& \left.\left.2N_c\left(
N_c-nN_f\right) \frac{m^2}{S^2_0}\right\} +
O(\Delta^3)\right],\\\langle
Z\rangle&=&Z_0\left[1-\Delta\left\{\frac{
n^2N_f\left(N_c-N_f\right)\left(2N_c-N_f\right)}{2N^2_c}\right\}+O(\Delta^2)
\right].\nn\eq This agrees with the results of  Ref.
\cite{Binetruy:1996uv}, except that there are two apparent typos
in Eq. (2.22) of that paper.

Now the $F$ and the $D$--terms are given by
\begin{eqnarray}\label{deltaF}
\langle
F_S\rangle&=&mS_0\Delta\left(nN_f\right)\left(1+\Delta\frac{
nN_f}{2}\left(n-1+\frac{nN_f\left(N_c-N_f\right)\left(2N_c-N_f\right)}{N_c^2}\right)\right),\nn\\
\langle F_Z\rangle&=&mZ_0\Delta\left(n^2N_f\right)
\left(\frac{N_f}{N_c}-1\right),\\
\langle D_A\rangle&=&m^2\Delta^2\left(n
N_f\right)^2\left(\frac{nN_f}{N_c}-1\right)/g_A\nn.
\end{eqnarray}
Consequently, the scalar soft masses induced from the $D$--term of
anomalous $U(1)$ are \bq\label{softM0}
m^2_{\tilde{f}_i}=q_{f_i}m_0^2,\eq where \bq\label{defm0}
m_0^2=m^2\Delta^2\left(nN_f\right)^2\left(\frac{nN_f}{N_c}-1\right).\eq
There is a simple interpretation of these results in terms of the
gaugino condensate (for $N_f < N_c-1$), which is given by
\cite{int}
\begin{equation}
\left\langle \lambda_\alpha \lambda^\alpha \right\rangle =
e^{2i\pi k/(N_c-N_f)} \Lambda^3 \left({m \over
\Lambda}\right)^{N_f/N_c},~~~ k=1 - (N_c-N_f).
\end{equation}
The soft scalar masses are simply proportional to the gaugino
condensate. We will make use of these results in Sec. 3. Note that
had we chosen $n=1$ these results would have led to negative
squared masses for scalars.  Note also that the $D$--term
contributions are proportional to the $U(1)_A$ charges, so they
are zero for particles with zero charge.

\subsection{Gravity corrections to the soft parameters}

~~~~~In this section we work out the supergravity corrections to
the soft parameters found in the global SUSY limit in the previous
section.  Our reasons for this extension  are two--fold.  First,
we wish to show explicitly that supergravity corrections do not
destabilize the minimum of the potential that we found in the
global limit.  Second, the main contribution to the masses of
scalars with zero $U(1)$ charge will arise from supergravity
corrections.  In our explicit models, we do have particles with
zero charge.

It is conventional in supergravity to add a constant term to the
superpotential in order to fine--tune the cosmological constant to
zero:
\begin{eqnarray}\label{superP2}
W=W_{global}+\beta.
\end{eqnarray}
We separate the constant into two parts, $\beta=\beta_0+\beta_1$,
such that $\beta_0$ cancels the leading part of the superpotential
in which case  $\langle W\rangle=\beta_1$. The $F$--term
contribution to the scalar potential in supergravity is given by
\begin{eqnarray}\label{scalarPotential1}
V_F=M_{Pl}^4e^G\left(G_i\left(G^{-1}\right)^i_jG^j-3\right),
\end{eqnarray}
where
\begin{eqnarray}\label{KahlerPotential}
G^i\equiv \partial G/\partial \phi_i^*,\,\,\,G_i\equiv \partial
G/\partial \phi^i,\,\,\,G^i_j\equiv \partial^2 G/\partial
\phi_i^*\partial \phi^j.
\end{eqnarray}
We will assume for illustration the minimal form of the K\"{a}hler
potential.  In our model it is given by
\begin{eqnarray}\label{KahlerPotential1}
G=\frac{|S|^2}{M_{Pl}^2}+2\frac{\mbox{Tr}(Z^\dagger
Z)^{1/2}}{M_{Pl}^2}+\sum_i\frac{|\phi_i|^2}{M_{Pl}}+
\mbox{ln}\left(\frac{|W|^2}{M_{Pl}^6}\right).
\end{eqnarray} Then the scalar potential is given by
\bq\label{scalarPotential2}V=V_F+V_D,\eq with
\bq\label{VFVD}V_F&=&e^{\left(|S|^2+2\mbox{Tr}(Z^\dagger
Z)^{1/2}+\sum_i|\phi_i|^2\right)/M_{Pl}^2}\left(\left|F_S+S^{*}\frac{W}{M_{Pl}^2}
\right|^2\right.\\&+&\left. {1 \over
2}\mbox{Tr}\left[\left(F_Z^\dagger+Z\frac{W}{M_{Pl}^2}\right)\left(Z^\dagger
Z\right)^{-1/2}\left(F_Z+Z^\dagger\frac{W}{M_{Pl}^2}\right)\right]\right.\\&+&\left.
\sum_i\left|F_{\phi_i}+\phi_i^*\frac{W}{M_{Pl}^2}\right|^2\right)-3\frac{\left|W\right|^2}{M_{Pl}^2},\eq
and \bq\label{VD}
V_D=\frac{g^2}{2}\left(G_i\left(T_a\right)^i_j\phi^j\right)^2M_{Pl}^4.\eq
In our case for $G^i=\phi^i/M_{Pl}^2+\partial W/\partial
\phi_i^*/W$, so $ M_{Pl}^2G_i\left(T_a\right)^{i}_j\phi^j=
\phi_i^*\left(T_a\right)^i_j\phi^j$, which is identical to the $D$
term of global supersymmetry (Note that the term $\partial
W/\partial \phi^i(T_a)^i_j\phi^j$ vanishes due to the gauge
invariance of $W$).

Including these supergravity corrections, by minimizing the
potential we find
\begin{eqnarray}\label{deltas1}\langle
S^\dagger S\rangle&=&\langle S^\dagger
S\rangle_{global}+2\Delta^2S^2_0\e^2\left[-\frac{n^2N_f^2}{4g_A^2N_c^2}
\left\{ng_A^2\left(N_c-N_f\right)^2+2N_c\left(nN_f+N_c\right)\frac{m^2}{S^2_0}\right\}\right.\nn
\\&-&\left.\frac{\tilde{\beta}_1nN_f}{4g_A^2N_c^2}\left\{g_A^2\left(N_c-N_f\right)^2
\left(2N_c+n\left(N_c-N_f\right)\right)+2N_c\left(nN_f-4N_c\right)\frac{m^2}{S^2_0}\right\}\right],\\
\langle Z\rangle&=&\langle Z\rangle_{global}-\Delta
\left(Z_0\e^2\right)\frac{N_c-N_f}{2N_c^2}\left[n^2N_f\left(N_c-N_f\right)
+\tilde{\beta}_1\left\{2N_c+n\left(N_c-N_f\right)\right\}\right],\nn\end{eqnarray}
where the subscript ``global" denotes the contributions found in
global SUSY case in Eq. (\ref{deltas}). Here we have introduced a
dimensionless parameter $\tilde{\beta}_1$ defined through the
relation \bq\label{defbeta}\beta_1=\left(\tilde{\beta}_1
mS^2_0\right)\Delta.\eq

From the condition that the vacuum energy is zero at the minimum
for the vanishing of the cosmological constant, $\tilde{\beta}_1$
is found to be \bq\label{beta}\tilde{\beta}_1\simeq
\pm\frac{nN_f}{\sqrt{3}\e} \left(1\pm
\frac{\e}{\sqrt{3}}+\frac{2}{3}\e^2\right).\eq Eq. (\ref{defbeta})
ensures that the cosmological constant remains zero to the scale
of strong dynamics. With these corrections the soft scalar masses
from the $D$--term are now given by
\bq\label{softMs2g}m^2_{\tilde{f}}=\left(m^2_{\tilde{f}}\right)_{global}
+q_fm^2_0\frac{\e^2
}{nN_f-N_c}\left[N_c+nN_f+\tilde{\beta}_1\left(1-4N_c/(nN_f)\right)\right].\eq
Note that the shifts in the masses are small, suppressed by a
factor of $\epsilon \simeq 0.2$.

The gravitino mass is determined to be
\bq\label{gravitinoM}m_{3/2}\simeq m\frac{\tilde{\beta}_1 \Delta
S_0^2}{M_{Pl}^2}\simeq \frac{
nN_f\Lambda^3}{\sqrt{3}S_0M_{Pl}}\left(\frac{m}{\Lambda}\right)^{N_f/N_c}.\eq
In addition to the $D$--term corrections, all scalar fields
receive a contribution to their soft masses from the term
\bq\label{softmq0}\left|\phi_i^*\frac{W}{M_{Pl}^2}\right|^2=
m_{3/2}^2|\phi_i|^2.\eq For particles neutral under the anomalous
$U(1)_A$ these are the leading source for soft masses.  With the
assumed minimal  K\"{a}hler potential, note that these soft masses
are equal to the gravitino mass.

So far we assumed the minimal form of the K\"{a}hler potential for
illustration.  There is no justification for this assumption. In
fact, within Split Supersymmetry, since there are no excessive
FCNC processes, an arbitrary form for the  K\"{a}hler potential is
permissible phenomenologically.  The effects of such a nonminimal
$G$ can be understood in terms of higher dimensional operators
suppressed by the Planck scale.  Scalar fields can acquire soft
SUSY breaking masses through the terms
\begin{equation}
\mathcal{L}\supset \int (\phi^*_i\phi^i){{|S|^2 \over M_{Pl}^2}}
d^4\theta~.\end{equation} The resulting masses are
$m_{\tilde{f_i}}^2 =c_i  m_{3/2}^2$, with $c_i$ being order one
(flavor--dependent) coefficients.  We will allow for such
corrections.

\section{Explicit models}

~~~~~In this section we consider a class of models based on
flavor--dependent anomalous $U(1)$ symmetry and apply the results
of the previous section.  These models were developed to address
the pattern of fermion masses and mixings
\cite{IbanezU1,Babu:2003zz}. As noted earlier, the anomalous
$U(1)$ $D$--term provides a small expansion parameter $\epsilon =
\left\langle S \right\rangle/M_{Pl} \sim 0.2$, which can be used
to explain the mass hierarchy.  We assign charge $q_i$ to fermion
$f_i$ and charge $q_j^c$ to fermion $f^c_j$, such that the mass
term $f_i f^c_j H$ will arise through a higher dimensional
operators with the factor $(S/M_{Pl})^{q_i+q_j^c}$ and thus
suppressed by a factor $\epsilon^{q_i+q_j^c}$. By choosing the
charges appropriately the observed mass and mixing hierarchy can
be explained, even with all Yukawa coefficients being of order
one.

With sub--TeV supersymmetry this approach to fermion mass and
mixing hierarchy cannot be combined with supersymmetry breaking
triggered by anomalous $U(1)$, since the $D$--terms will split the
masses of scalars leading to unacceptable FCNC.  Within Split
Supersymmetry, however, these two approaches can be combined,
which is what we analyze now.

The superpotential of the class of models under discussion has the
following form:

\begin{eqnarray}\label{superP1}
W&=&\sum_{f}y_{ij}^f  f_iHf^c_j
\left(\frac{S}{M_{Pl}}\right)^{n^f_{ij}}+{{M_{R}}_{ij} \over 2}
\nu^c_i \nu^c_j\left(\frac{S}{M_{Pl}}\right)^{n^{\nu^c}_{ij}}+\mu
H_u H_d\nn\\&+&\frac{\mbox{Tr}\left(\lambda
Z\right)S^n}{M_{Pl}^{n-1}}+(N_c-N_f)\left(\frac{\Lambda^{3N_c-N_f}}
{\mbox{det}\left(Z\right)}\right)^{1/(N_c-N_f)}+W_{A}\left(S,
X_k\right)\, .
\end{eqnarray}
Here $X_k$ are the SM singlet fields necessary for the cancelation
of gravitaitonal anomaly.  We will focus on the sub-class  of such
models studied in Ref. \cite{Babu:2003zz}.  The mass matrices for
the various sectors  in Ref. \cite{Babu:2003zz} are given (in an
obvious notation) by:

\begin{eqnarray}\label{massM1}
&&M_u\sim \langle
H_u\rangle\pmatrix{\epsilon^{\,8-2\alpha}&\epsilon^{\,6-\alpha}&\epsilon^{\,4-\alpha}\cr
\epsilon^{\,6-\alpha}&\epsilon^4&\epsilon^2\cr\epsilon^{\,4-\alpha}&\epsilon^2&1}\,,\hspace{1.cm}
M_d\sim \langle H_d\rangle\epsilon^p
\pmatrix{\epsilon^{\,5-\alpha}&\epsilon^{\,4-\alpha}&\epsilon^{\,4-\alpha}\cr
\epsilon^3 &\epsilon^2&\epsilon^2\cr\epsilon&1&1},\nn\\
\nn\\
\vspace{.5cm} &&M_e\sim \langle
H_d\rangle\epsilon^p\pmatrix{\epsilon^{\,5-\alpha}&\epsilon^3&\epsilon\cr
\epsilon^{\,4-\alpha}&\epsilon^2&1\cr\epsilon^{\,4-\alpha}&\epsilon^2&1}\,,\hspace{1.cm}
M_{\nu_D}\sim \langle H_u\rangle\epsilon^p
\pmatrix{\epsilon^2&\epsilon &\epsilon \cr \epsilon
&1&1\cr\epsilon&1&1},\nn\\
\nn\\
 \vspace{.5cm} &&M_{\nu^c}\sim M_R
\pmatrix{\epsilon^2&\epsilon&\epsilon\cr
\epsilon&1&1\cr\epsilon&1&1}\,\hspace{.5cm}\Rightarrow
\hspace{.5cm} M^{light}_\nu\sim \frac{{\langle
H_u\rangle}^2}{M_R}\epsilon^{2p}
\pmatrix{\epsilon^2&\epsilon&\epsilon\cr
\epsilon&1&1\cr\epsilon&1&1}.
\end{eqnarray}
Although not unique, these mass matrices would lead to small quark
mixings and large neutrino mixings. Note that the neutrino masses
are hierarchical in this scheme.

The charge assignment which leads to these mass matrices is given
in Table \ref{table1}. Here we use $SU(5)$ notation  for the
fields in the first column for simplicity, although we do not
explicitly assume $SU(5)$ unification.  There are two parameters,
$p$ and $\alpha$, which can take a set of discrete values. The
parameter $p$ takes values  $p=2\,~(1,\,0)$ corresponding to low
(medium, high) value of $\tan\beta$ (the ratio of the two Higgs
VEVs). Actually, in Split Supersymmetry, since $\tan\beta \sim 1$
is also permitted, $p=3$ is also allowed.  $\alpha$ appears in the
mass of the up--quark, both $\alpha=0$ and $\alpha=1$ give
reasonable spectrum.  We also consider the case where the charge
of $\bar{5}_1$ is $p$ (rather than $1+p$) in Table 1.  This case
would have mass matrices which are very similar to those in Eq.
(42). The main difference in this case is that all elements of
$M_{\nu}^{light}$ will be of the same order, which would lead to
larger $U_{e3}$. This scenario has been widely studied
\cite{Babu:1995hr}, sometimes under the name of neutrino mass
anarchy \cite{Hall:1999sn}.
\begin{table}[t]
\begin{center}
\begin{tabular}{|c|c|}\hline
\rule[5mm]{0mm}{0pt}Field& Anomalous flavor charges\\\hline
\rule[5mm]{0mm}{0pt}$10_1,\, 10_2,\, 10_3$& \,\,$ 4-\alpha, \,2, \,0$\\
\rule[5mm]{0mm}{0pt}$\overline{5}_1,\, \overline{5}_2,\, \overline{5}_3$&\,$ 1+p,\, p,\,p$\\
\rule[5mm]{0mm}{0pt}$\nu^c_1,\,\nu^c_2,\,\nu^c_3$ &$ 1,\, 0,\, 0$\\
\rule[5mm]{0mm}{0pt}$H_u$, $H_d$, $S$, $Q$, $\tilde{Q}$ & $ 0,\, 0,\, -1,\, n/2$\\
\hline
\end{tabular}
\caption{\footnotesize The flavor $U(1)_A$ charge assignment for
the MSSM fields, the $SU(N_c)$ fields $Q$ and $\tilde{Q}$ and the
flavon field $S$ in the normalization where $q_S=-1$.}
  \label{table1}
\end{center}
\end{table}
The charge assignment of Table 1, as well as its above--mentioned
variant, explain naturally the mass and mixing hierarchy of quarks
and leptons, including small quark mixings and large neutrino
mixings.

The Green--Schwarz anomaly cancelation conditions for these models
are given by \bq\label{anomaly2} {A_1 \over k_1} ={A_i \over k_i}
= \frac{A_{N_c}}{k_N}=\frac{nN_f}{2k_N}=\frac{19-3\alpha+3p}{2k_i}
\,\,\,\mbox{or}\,\,\,\frac{18-3\alpha+3p}{2k_i}\eq with $A_i$
being the $(SM)^2 \times U(1)_A$ anomalies for $i=2-3$. Their
equality is  automatically satisfied, due to the $SU(5)$
compatibility of charges, provided that the Kac--Moody levels
$k_i$ for the SM gauge groups $U(1)_Y$, $SU(2)_L$ and $SU(3)_c$
are chosen to be, for example, $5/3,\,1$ and $1$ respectively. For
$A_{\rm gravity}$, one needs to introduce extra heavy matter $X_k$
(with charge $+1$)  which decouple at or near the Planck scale
(see Ref. \cite{Babu:2003zz} for a detailed discussion). In Eq.
(43) the first $p$--dependent factor applies to the charge
assignment of Table 1, while the second one corresponds to the
variant with $\bar{5}_1$ carrying charge $p$.  For every choice of
charge we can compute the expansion parameter $\epsilon$ from
$\label{eps}\e=\sqrt{g_{st}^2A_{\rm gravity}/(192\pi^2)}$.
 We find for $\alpha=0$ and for the charges of Table 1,
 $\epsilon =0.174\,(0.187,\,0.199)$ for $p=0\,(1,\,2)$.
The results are very similar for other choices.

Eq. (\ref{anomaly2}) allows for only a finite set of choices for
$n$,  $N_c$ and $N_f$.  First of all, all these must be integers.
Secondly, the mass parameter $m$ of the meson fields of $SU(N_c)$
must be of order $\Lambda$ or smaller, otherwise these mesons will
decouple from the low energy theory, affecting its dynamics.
Thirdly, the dynamical scale $\Lambda$ is determined for any
choice of charges, due to the string unification condition, Eq.
(3).  (We will confine to Kac--Moody level 1 for the $SU(N_c)$ as
well as the SM sectors.)  This should lead to an acceptable SUSY
breaking spectrum. Consistent with these demands, we find four
promising cases.  (i) $n=5,~N_f = 5,~p=2,~\alpha=0$;  (ii)
$n=6,~N_f=4,~p=2,~\alpha=0$; (iii) $n=7,~N_f=3,~p=1,~\alpha=1$;
and (iv) $n=6,~N_f=3,~p=1,~\alpha=1$.  Here (i) has $\bar{5}_1$
charge equal to $p+1$, while the other three cases has it to be
equal to $p$. We will see that the choices $N_c=6$ or $7$ yield
reasonable spectrum.

\subsection{The spectrum of the model}

~~~~~Now we turn to the spectrum of the model. We set the gaugino
masses at the TeV scale.  (The Higgsinos will turn out to have
masses of the same order.)  We then seek possible values of the
scale $\Lambda$ and the  mass parameter $m_0$ (the scalar mass)
that would induce the TeV scale gaugino masses. The spectrum will
turn out to be that of Split Supersymmetry.  The main reason for
this is that the leading SUSY breaking term, the
$U(1)_A$~$D$--term, generates squark and slepton masses, but not
gaugino and Higgsino masses.

Supersymmetry breaking trilinear $A$ terms are induced in the
model by the same superpotential $W$ (Eq. (41)) that generates
quark and lepton masses, once the $S$ field acquires a nonzero $F$
component: \bq\label{Aterm} \mathcal{L}\supset y^f_{ij}\int
d^2\theta f_if^c_jH \left(\frac{S}{M_{Pl}}\right)^{n^f_{ij}} =
Y^f_{ij}\left(q^f_i+q^{f^c}_j\right)\tilde{f}_i\tilde{f}^c_jH\frac{F_S}{S}.\eq
Here $Y^f_{ij}\simeq y^f_{ij}\e^{n^f_{ij}}$ are the effective MSSM
Yukawa couplings, with $n_{ij}= q_{f_i}+q_{f_j}^c$, the sum of the
anomalous charge of the SM fermions $f_i$ and $f^c_j$.
Substituting results from the previous section, Eqs. (20) and
(21), we find \bq\label{Aterm1} A^f_{ij}=
Y^f_{ij}\left(q^f_i+q^{f^c}_j\right)nN_f\frac{\Lambda^3}{S_0^2}\left(
\frac{m}{\Lambda}\right)^{N_f/N_c}.\eq

These $A$--terms are induced at the scale $\Lambda$.  The
messengers of supersymmetry breaking are the meson fields of the
$SU(N_c)$ sector, which have masses of order $\Lambda$.  In the
momentum range $m_0 \leq \mu \leq \Lambda$, the spectrum is that
of the MSSM and there is renormalization group running of all SUSY
breaking parameters as per the MSSM beta functions. This implies
that once the $A$--terms are induced, they will generate nonzero
gaugino masses through two--loop MSSM interactions. These are
estimated from the two--loop MSSM beta functions to
be\footnote{The one--loop finite corrections arising from diagrams
involving the top--quark and the stop--squark are negligible since
$A_t=0$ and $\mu\sim$ TeV $\ll\,m_{\tilde{t}}.$ }
\bq\label{gauginoM0}M^i_{\tilde{g}}(m_0)&\simeq&-\frac{g_i^2}{(16\pi^2)^2}
\left(C^b_iY^2_b+C^\tau_iY^2_\tau\right)\frac{m_0}{\sqrt{nN_f/N_c-1}}
\mbox{ln}\left(\Lambda^2/m_0^2\right),\eq where
$C^b_i=\left(14/5,6,4\right)$ and $C^\tau_i=\left(18/5,6,0\right)$
for $i=1-3$. $Y_b$ and $Y_\tau$ are the MSSM Yukawa couplings of
the $b$--quark and the $\tau$--lepton. From the requirement that
$M^i_{\tilde{g}}\sim 1$ TeV we can estimate $\Lambda$ and $m_0$,
which will enable us to obtain the full spectrum of the model.
Assuming that $m\sim\Lambda$, for the Bino mass we obtain (for
$p=2$, or $\tan\beta \sim 5$):
\bq\label{nmB2l}M_{\tilde{B}}(m_0)&\sim&-10^{-5}m_0.\eq The mass
of the Wino is somewhat larger than this, and that of the gluino
is somewhat smaller (compare the coefficients $C_i^b$ and
$C_i^\tau$), all at the scale $m_0$.  There is significant running
of these masses below $m_0$ down to the TeV scale.  This running
is the largest for the gluino \cite{gr} which increases its mass,
while it is the smallest for the Bino, which decreases its mass.
Consequently, at the TeV scale, we have the normal mass hierarchy
$M_{\rm Bino} \leq M_{\rm Wino} \leq M_{\rm gluino}$.

In addition to the SM gauge interactions, the gauginos receive
masses from the anomaly mediated contributions \cite{amsb}.  These
contributions may be suppressed in specific setups such as in 5
dimensional supergravity \cite{Arkani-Hamed:2004fb}.  We will
allow for both a suppressed and an unsuppressed anomaly mediated
contributions to gaugino masses.  These contributions are given by
\begin{equation}
M_{\rm gaugino} = {\beta(g) \over g} F_\phi
\end{equation}
where $F_\phi$ is the $F$--component of the compensator
superfield. With our setup as described in the previous section,
$F_\phi$ is equal to the gravitino mass, so the Wino mass, for
eg., will be about $3 \times 10^{-3}$ of the gravitino mass, or
about $10^{-3} m_0$.  If we set the Wino mass at 1 TeV, $m_0$ will
be of order $10^6$ GeV in such a scenario.

As we stated in the previous section, only a limited choice of $n$
and $N_f$ are allowed from the mixed anomaly cancelation
conditions. We have considered four cases with $n N_f = 25, ~24,
~21,$ or $18$. Our results for the spectrum are listed in Table 2.
In each case we studied different values of $N_c > N_f$.  $N_c =
6,~7$ give the correct dynamical scale $\Lambda$ which leads to
TeV scale gauginos.  The scalar masses are found to be of order
$10^6$ GeV in the case of unsuppressed anomaly mediated
contribution (cases 1 and 3), and of order $10^8$ GeV for the
suppressed case (all the other cases). Clearly this is a  Split
Supersymmetry spectrum. In the computation of Table 2 we assumed
$g_{N_c}^2/(4\pi)=1/28$ at $M_{Pl} = 2.4 \times 10^{18}$ GeV.  The
mass $m$ for the meson fields is computed in terms of an effective
coupling $\hat{\lambda}\equiv\lambda
\left(\frac{M_{Pl}}{M_*}\right)^{n-1}$.  We expect $\hat{\lambda}$
to be of order one from naturalness, if $M_*$ is the same as
$M_{Pl}$.  We list the mass $m$ in terms of $\hat{\lambda}$ in the
third column in Table 2. Note that the scalar masses from
anomalous $U(1)$ ~$D$--term are proportional to the $U(1)$
charges, and therefore vanish for $H_u,~H_d$ and $10_3$ fields.
These fields will however acquire masses from supergravity
corrections.

\begin{table}[t]
\begin{center}
\begin{tabular}{|c|c|c|c|c|c|} \hline
\rule[5mm]{0mm}{0pt}$(p,\,\alpha,\,n,\,N_f,\,N_c)$&$
\Lambda\left(\mbox{GeV}\right)$ &$
m\left(\mbox{GeV}\right)/\hat{\lambda}$&\,$
m_0$&$M_{\tilde{B}}\left({m_0}\right)\,\left(\mbox{GeV}\right)$&\,
$\mu\left(\mbox{GeV}\right)/\lambda_\mu$\\\hline
\rule[5mm]{0mm}{0pt}$\left(2,\,0,\,5,\,5,\,6\right)$&$3\times10^{12}$
&$8\times10^{14}$
&$6\times10^{5}\hat{\lambda}^{5/6}$&$5\hat{\lambda}^{5/6}$&$600/\hat{\lambda}^{1/6}$\\\hline
\rule[5mm]{0mm}{0pt}$\left(2,\,0,\,5,\,5,\,7\right)$&$4\times10^{13}$&$8\times10^{14}$
&$9\times10^{7}\hat{\lambda}^{5/7}$&$600\hat{\lambda}^{5/7}$&$9\times10^4/\hat{\lambda}^{2/7}$\\\hline
\rule[5mm]{0mm}{0pt}$\left(2,\,0,\,6,\,4,\,6\right)$&$8\times10^{12}$&$1\times10^{14}$
&$7\times10^{5}\hat{\lambda}^{2/3}$&$5\hat{\lambda}^{2/3}$&$700/\hat{\lambda}^{1/3}$\\\hline
\rule[5mm]{0mm}{0pt}$\left(2,\,0,\,6,\,4,\,7\right)$&$8\times10^{12}$&$1\times10^{14}$
&$1\times10^{8}\hat{\lambda}^{4/7}$&$640\hat{\lambda}^{4/7}$&$10^5/\hat{\lambda}^{3/7}$\\\hline
\rule[5mm]{0mm}{0pt}$\left(1,\,0,\,7,\,3,\,6\right)$&$2\times10^{13}$&$3\times10^{13}$
&$1\times10^{6}\hat{\lambda}^{1/2}$&$100\hat{\lambda}^{1/2}$&$1600/\hat{\lambda}^{1/2}$\\\hline
\rule[5mm]{0mm}{0pt}$\left(1,\,0,\,7,\,3,\,7\right)$&$1\times10^{14}$&$3\times10^{13}$
&$2\times10^{8}\hat{\lambda}^{3/7}$&$10^4\hat{\lambda}^{3/7}$&$2\times10^5/\hat{\lambda}^{4/7}$\\\hline
\rule[5mm]{0mm}{0pt}$\left(1,\,1,\,6,\,3,\,6\right)$&$2\times10^{13}$&$1\times10^{14}$
&$2\times10^{6}\hat{\lambda}^{1/2}$&$200\hat{\lambda}^{1/2}$&$3000/\hat{\lambda}^{1/2}$\\
\hline
\end{tabular}
\caption{\footnotesize The spectrum of the model for different
choices of $p,~\alpha,~n,~N_f$ and $N_c$.  In computing $\Lambda$,
we use Eq. (10) with $\alpha_{N_c} = 1/28$ at the Planck scale.
The Bino mass estimate is very rough, and includes only the
two--loop MSSM induced contributions.}
  \label{table2}
  \end{center}
\end{table}

The $U(1)_A$ symmetry does not forbid a bare $\mu$ term in the
superpotential.  However, it can be banished by a discrete $Z_4$
$R$--symmetry \cite{bgw}.  Under this $Z_4$,  all the SM fermion
superfields (scalar components) have charge $+1$, the gauginos
have charge $+1$, the $Z$ field has charge $+2$ and the SM Higsses
and the $S$ fields have charge zero.  This symmetry has no
anomaly, as a consequence of discrete Green--Schwarz anomaly
cancelation.  The $G^2 \times Z_4$ anomaly coefficients are $A_3 =
3,~A_2 = 2-1 = 1$ and $A_{N_c} = N_c$.  The GS condition for
discrete $Z_4$ anomaly cancelation is that the differences $A_i -
A_j$ should be an integral multiple of 2, which is automatic when
$N_c$ is odd.

One can write the following effective Lagrangian for the $\mu$
term that is consistent with the $Z_4$~ $R$ symmetry:
\bq\label{mu1} \mathcal{L}\supset\int d^2\theta H_uH_d
\frac{\mbox{Tr}(\lambda_\mu Z)S^n}{M_{Pl}^{n+1}} = \lambda_\mu N_f
\frac{\langle ZS^n\rangle}{M^{n+1}_{Pl}}H_uH_d.\eq This leads  to
\bq\label{mu2}\mu=\lambda_{\mu}\e^nN_f\frac{\Lambda^3}{mM_{Pl}}
\left(\frac{m}{\Lambda}\right)^{N_f/N_c}.\eq The numerical results
for $\mu$--term are given in the last column of Table \ref{table2}
using this relation.

The SUSY breaking bilinear Higgs coupling, the $B\mu$ term, arises
from the Lagrangian \bq\label{Mmu1} L \supset &=&\int d^4\theta
H_uH_d\frac{\lambda_{B1}|S|^2+\lambda_{B2}\mbox{Tr}\left(Z^\dagger
Z\right)^{1/2}}{M_{Pl}^2}\nn\\&=&
\frac{\lambda_{B1}|F^0_S|^2+\lambda_{B2}N_f|F_Z|^2/|Z_0|}{M_{Pl}^2}H_uH_d,\eq
leading to \bq\label{Bmu1}
B\mu&=&m^2_0\left(\frac{N_c}{nN_f-N_c}\right)^2\left(\lambda_{B1}\e^2
+\lambda_{B2}n^2N_f\frac{\Lambda^3}{mM_{Pl}^2}
\left(\frac{m}{\Lambda}\right)^{N_f/N_c}\right).\eq The second
term in Eq. (\ref{Bmu1}) is small compared to the first. From this
we see that the $2 \times 2$ Higgs boson mass matrix has its
off--diagonal entry of the same order as its diagonal entries.
Recall that the diagonal entries are of order $m^2_{3/2}$, since
the $U(1)_A$ charges of $H_u$ and $H_d$ are zero. Fine--tuning can
then be done consistently so that one of the Higgs doublets remain
light, with mass of order $10^2$ GeV.

Even when the $Z_4$ $R$ symmetry is not respected by gravitational
corrections, the induced $\mu$ term and gaugino masses are of
order TeV.  There can be a new contribution to the $\mu$ term in
this case, arising from
\begin{equation}
L \supset \int d^4 \theta H_u H_d {(ZS^n)^* \over M_{Pl}^{n+2}}.
\end{equation}
This $\mu$ term is however smaller than that from Eq. (49).
Similarly, gaugino masses can arise from
\begin{equation}
L \supset \int d^4 \theta W_\alpha W^\alpha {ZS^n \over
M_{Pl}^{n+2}}
\end{equation}
which is also smaller than the SM induced corrections.

 For the scalars neutral under $U(1)_A$ ($H_u, ~H_d$
and $10_3$), the $D$--term contribution to the soft masses vanish.
We should take account of the subleading supergravity corrections
then. Since these corrections are suppressed by a factor of
$\epsilon^2$ in the mass--squared, we should worry about
potentially large negative corrections proportional to the other
soft masses arising from SM interactions through the RGE in the
momentum range $m_0 \leq \mu \leq \Lambda$. We have examined this
in detail and found consistency of the models.

For the masses of zero charge fields we write
\bq\label{massq0}m^2_{\tilde{\phi_i}}=c_i
m^2_{3/2}+\delta\left(m^2_{\tilde{\phi_i}}\right)\eq with
$\delta\left(m^2_{\tilde{\phi_i}}\right)$ denoting the MSSM RGE
corrections.  The most prominent one--loop radiative corrections
are \bq\label{ms1loop}\delta
\left(m^2_{\tilde{f}_3}\right)^{1-loop}&\simeq&-\left(Y_b\right)^2\frac{m^2_0}{16\pi^2}
\frac{p}{\left(nN_f/N_c-1\right)}\frac{nN_f}{N_c}\mbox{ln}
\left(\frac{\Lambda^2}{m^2_{\tilde{f}}}\right),\nn\\
\delta \left(m^2_{H_d}\right)^{1-loop} &\simeq& -\left\{3(Y_b)^2 +
(Y_\tau)^2 \right\}\frac{m^2_0}{16\pi^2}
\frac{p}{\left(nN_f/N_c-1\right)}\frac{nN_f}{N_c}\mbox{ln}
\left(\frac{\Lambda^2}{m^2_{\tilde{f}}}\right)\eq where
$\tilde{f}_3=(\tilde{Q}_3,\,\tilde{e_3^c})$. Similar corrections
for $H_u$ and $\tilde{u_3^c}$ scalar components are small. Since
$p=2$, we have low $\tan\beta \sim 5$, so these corrections are
not large, although not negligible. For example, for the
down--type Higgs bosons we have
\bq\label{radcorr1}\delta\left(m^2_{H_d}\right)^{1-loop}\sim
-2\times10^{-3}m^2_0.\eq  If the supergravity corrections to the
mass--squared of $H_d$ is larger than $3 \times 10^{-2} m_0$, it
will remain positive down to the scale $m_0$.

There is an important two--loop correction to the scalar masses
arising from the gauge sector:
 \bq\label{ms2loop}\delta
\left(m^2_{\tilde{\phi}}\right)^{2-loop}&\simeq&-g^4\frac{m^2_0}{(16\pi^2)^2}(nN_f)
K_\phi\mbox{ln} \left(\frac{\Lambda^2}{m_{\tilde{f}}^2}\right)\eq
where $K_\phi = (63/15 ,\, 16/5, \,6/5\,\,\mbox{and}\,\,\,9/5)$
for
$\tilde{\phi}=(\tilde{Q},\,\tilde{u^c},\,\tilde{e^c}\,\,\mbox{and}\,\,\,H_u)$.
This correction is estimated to be
\bq\label{radcorr2}\left(m^2_{\tilde{f}}\right)^{2-loop}\sim-10^{-2}m_0^2.\eq
We see that these corrections are, although close to the gravitino
contribution, at a safe level.  We conclude that Split
Supersymmetry is realized consistently in these models.

\section{Conclusion}

~~~~~In this paper we have proposed concrete models for
supersymmetry breaking making use of the anomalous $U(1)$
~$D$--term of string origin.  The anomalous $U(1)$ sector is
coupled to the strong dynamics of an $N=1$ SUSY gauge theory where
exact results are known.  The complete models we have presented
also address the mass and mixing hierarchy of quarks and leptons.
We have generalized the analysis of Ref. \cite{Binetruy:1996uv} to
include supergravity corrections, which turns out to be important
for certain fields in these models which carry zero $U(1)$ charge.
Table 2 summarizes our results on the spectrum of these models.
This spectrum is that of Split Supersymmetry.  The gaugino and the
Higgsino masses are of the same order, when these are set at the
TeV scale, the squarks and sleptons have masses in the range
$(10^6-10^8)$ GeV. This provides an explicit realization of part
of the parameter space of split supersymmetry
\cite{Arkani-Hamed:2004fb}.

The experimental and cosmological implications of Split
Supersymmetry have been widely studied
\cite{gr,other,Wells:2004di,nath}. We conclude by summarizing the
salient features that apply to our framework. (i) Gauge coupling
unification works well, in fact somewhat better than in the MSSM.
When embedded into $SU(5)$ symmetry, proton decay via dimension
six operators will result, with an estimated lifetime for $p
\rightarrow e^+ \pi^0$ of order $(10^{35}-10^{36})$ yrs.  There is
no observable $d=5$ proton decay in these models.  (ii) The
lightest neutralino, which is charge and color neutral, is a
natural and consistent dark matter candidate.  (iii) The gluino
lifetime is estimated to be of order $10^{-7}$ seconds or shorter
in these models.  There is no cosmological difficulty with such a
mass.  (iv) The gravitino mass is or order $10^7$ GeV, thus there
is no cosmological gravitino abundance problem. (v) The low energy
theory is the SM plus the neutralinos and the charginos of
supersymmetry. All other particles acquire masses either near the
Planck scale or through strong dynamics at a scale $\Lambda \sim
10^{14}$ GeV.

\section*{Acknowledgments}
~~~~~KB and TE are  supported in part by the US Department of
Energy under grants \#DE-FG02-04ER46140 and \# DE-FG02-04ER41306.
BM acknowledges hospitality of the Theory Group at Oklahoma State
University during a visit when this work started.

\end{document}